\documentclass[11pt]{article}

\usepackage{epsfig}
\usepackage{latexsym}
\usepackage{amssymb}

\newcommand{\sect}[1]{\setcounter{equation}{0}\section{#1}}

\def\be{\begin{equation}}
\def\ee{\end{equation}}
\def\ba{\begin{eqnarray}}
\def\ea{\end{eqnarray}}

\title{{\bf Zero-Branes, Quantum Mechanics and the Cosmological Constant}}
\author{{\bf A. Chamblin$^{1,2}$}\thanks{email: chamblin@mit.edu} 
\& {\bf N.D. Lambert$^{3}$}\thanks{email: lambert@mth.kcl.ac.uk}
\\ $^{1}$Center for Theoretical Physics, Massachusetts Institute of
Technology, Bldg. 6-304, 
\\ Cambridge, MA 02139, U.S.A.\\
$^{2}$Theory Division, T-8, Los Alamos National Laboratory, 
\\ Los Alamos, NM 87545, U.S.A.\\
 \\ $^{3}$Department of Mathematics, King's College, London, \\
 WC2R 2LS, United Kingdom.
\\ \\ LA-UR-01-2984
\\ \\ KCL-TH-01-22}

\begin{document}

\maketitle

\begin{abstract}
We analyse some dynamical issues in a 
modified type IIA supergravity, recently proposed as an extension of
M-theory that admits de Sitter space.
In particular we find that this theory has multiple zero-brane
solutions. This suggests a microscopic quantum mechanical 
matrix description which yields a massive deformation of the usual M(atrix)
formulation of M-theory and type IIA string theory.

\end{abstract}

\newpage
\section{Introduction}

Any compelling theory of cosmology must resolve the cosmological horizon,
flatness, and topological defect abundance problems, and must also generate
a scale-invariant spectrum of density fluctuations.  An elegant idea which
does all of these things is the idea of inflation, which asserts
that there was a period of accelerating expansion 
in the early universe.  The beauty of inflation is that it doesn't
depend on what happened before inflation, so it has many model-independent
properties. Furthermore recent observational results point towards a 
small but non-zero cosmological constant in our observable universe. 

It is now generally accepted that most fruitful candidate for a
unified description of the fundamental interactions, including quantum 
gravity,
is the mysterious `M-theory', which encompasses the perturbative 
superstring theories.  The low energy limit of this theory is 
eleven-dimensional supergravity, and compactifications of this 
theory on circles, tori and orbifolds give rise to the perturbative
string  theories in ten dimensions.  While the M-theory moduli space 
contains a huge variety
of supergravity solutions, to date it has proved remarkably difficult to 
obtain inflation or 
de Sitter space directly from compactification of the low-energy
limit of M-theory. In addition de Sitter space poses intriguing
difficulties for quantum theory since its  entropy is associated to a system
with only a finite number of degrees of freedom 
(e.g. see \cite{banks,witten}).

In a recent paper \cite{mm}, we have attempted to improve this situation.
In particular we studied a supergravity, dubbed `MM-theory',
which may be regarded as a slight extension of eleven-dimensional 
supergravity  
which still preserves eleven-dimensional supersymmetry. 
This extension describes  the  most general form
of eleven dimensional supersymmetry that is compatible with on-shell
superfields. This problem was studied in \cite{howe} and involves
extending the local structure group of spacetime from $Spin(1,10)$ to
a conformal spin group $CSpin(1,10)$. The superspace constriants then
assert that the conformal connection is locally trivial. In effect
this procedure gauges the global scale symmetries of the M-theory
equations of motion. 

Moreover, as long as the eleven-dimensional space is simply connected, 
this theory can be obtained through a gauge transformation of
the usual eleven-dimensional supergravity \cite{CJS}.
However if the manifold is not simply connected then it is possible
to introduce a Wilson line in the conformal connection. This is
analogous to the Higgs' effect in gauge theory compactified on a
torus. After compactification, the fields which are charged with respect
to the Wilson line then become massive in the
lower dimensional theory.

A crucial point is that it only really makes sense to physically
distinguish  `MM-theory' from M-theory if ${\pi}_{1}(M_{11}) \neq 0$.
Indeed, in compactified MM-theory one finds that 
the connection acts as a source for the Einstein tensor and the
natural vacuum is ten-dimensional de Sitter space. 

While this constuction shows that one can embed de Sitter space into
eleven-dimensional superspace and supergravity, its main 
drawback  is that it is not clear how, if at all,  
to relate MM-theory to
a microscopic quantum theory and in particular M-theory.
The equations of MM-theory compactified to ten-dimensions are
certainly rather odd. In addition they do not come from an action. 
However they have the advantage that they can be readily obtained from
the standard equations of M-theory. Indeed MM-theory provides a continuous
deformation of type IIA supergravity to include a positive
cosmological constant and hence de Sitter space. In addition 
the graviphoton has a tachyonic form at the linearised level 
(although it is not clear what is the case in the full non-linear theory). 
This should be contrasted with the usual analysis of the de Sitter
supergravities which admit actions but generally contain 
ghost-like vectors \cite{PvNS}.
The appearance of tachyons, rather than ghosts, is encouraging 
since they are much more easily understood within quantum theory.
Thus MM-theory allows us study the emergence of de Sitter space
from a deformation of type IIA supergravity and M-theory in a relatively
controlled way. Therefore we also hope that our work will be insightful for
other recent approaches to the study of de Sitter space from the string
theory point of view \cite{andy,li,klemm,vijay,gao,chull, moffat,gordon}.

Here we wish to address some dynamical issues and attempt to obtain an
underlying microscopic description of MM-theory. Our analysis suggests that 
MM-theory describes 
a legitimate part of M-theory, but  represents an unstable phase of the
theory. We will identify the existance of zero-branes which obey a 
no-force condition and thereby motivate a matrix model of the underlying
dynamics.

\section{MM-theory on $S^1$}

The Bosonic equations of MM-theory in eleven dimensions can be written as
\begin{eqnarray}
\label{elevenDact}
\hat R_{{ab}} - {1\over 2}\hat g_{{ab}}\hat R 
&=& -18\hat D_{ a} \hat k_{ b} 
+ 18 \hat g_{ {ab}}\hat D^{  c} \hat k_{ c}
 +36 \hat k_{ a}\hat k_{ b} + 144 \hat k^2 
\hat g_{ {ab}}
-{1\over 48}\left(4\hat H_{ {acbe}}
\hat H_{\underline b}^{\ \ {cde}} - \frac{1}{2}
g_{ {ab}}\hat H^2\right) \nonumber\\
\hat D^{ a}\hat H_{ {abcd}}
&=& -12\hat k^{ a}\hat H_{{abcd}}
+{1\over 36\cdot 48}\epsilon_{{bcde}...{ f}...}
\hat H^{{e}...}\hat H^{{f}...} \ , \nonumber\\
\end{eqnarray}
where $\hat H_{{abcd}}=4\partial_{[ a}
\hat B_{{bcd}]}+24k_{[a}\hat B_{{bcd}]}$ and $\partial_{[a}\hat k_{b]}=0$.
The superspace construction of these equations guarantees that the
the full Fermionic and Bosonic system is
invariant under the supersymmetry
\begin{eqnarray}
\label{susyone}
\delta \hat e_{a}^{\ \ \underline b} 
&=& -i\hat \epsilon\hat \Gamma^{\underline b}\ ,\nonumber\\
\delta \hat B_{abc} &=& -3i\epsilon\hat \Gamma_{[ab}\hat
\psi_{c]}\ ,\nonumber\\
\delta \hat\psi_a &=& \hat D_a \epsilon 
+\frac{1}{36}\left(\hat\Gamma^{bcd}\hat H_{abcd}
+\frac{1}{8}\hat\epsilon\hat\Gamma_{abcde}\hat H^{bcde}\right)\hat\epsilon
-\hat k^b \hat \Gamma_{ab}\hat\epsilon + 2\hat k_a\hat\epsilon
\ ,\nonumber\\ 
\end{eqnarray}
where $\hat\psi_a$ is the eleven dimensional gravitini.
Here we have written the  
Weyl superspace equations  of \cite{howe}  in terms of 
the standard eleven-dimensional Levi-Civita connection and curvature.
These equations are simply those of the
usual eleven-dimensional supergravity \cite{CJS} but with a $CSpin(1,10)$
connection \begin{equation}
\label{connection}
\hat \Upsilon_{a}=
\frac{1}{4}\left(\hat\omega_a^{\ \underline {bc}}
-4\hat e_a^{\ [\underline{b}}\hat k^{\underline{c}]}\right)
\hat\Gamma_{\underline {bc}} 
+2\hat k_a\ ,
\end{equation}
instead  of the usual $Spin(1,10)$
connection $\hat \omega_{a}^{\ \underline{bc}}$.

As discussed in \cite{neil}, when we perform a standard Kaluza-Klein
reduction of MM-theory on the $S^1$, we arrive at a {\it new} massive 
IIA supergravity, which has the appealing property that it can 
be obtained by compactification of 
eleven-dimensional supergravity on a circle, with 
the introduction of a `Wilson line' in the conformal connection.

The equations of motion for the low energy effective action of
MM-theory can be obtained by compactifying the equations of motion
\ref{elevenDact} with a Wilson line
in the Conformal connection, 
${\hat k}_{a}=m\delta_{a}^y$, 
and assuming that the
fields are independent of the eleventh-dimension $y$. These equations
were first constructed in \cite{neil}, although there the two-form and
three-form gauge fields were set to zero. Here we follow the
methods and conventions of \cite{neil}, but include all the Bosonic
fields. In particular the eleven-dimensional vielbein has the form
\begin{equation}
\label{veilbein}
{\hat e}_{a}^{\ \ \underline b} = \left(
\matrix{e^{-\phi/3}e_m^{\ \ \underline n}&A_me^{2\phi/3}\cr
0&e^{2\phi/3}}\right)\ .
\end{equation}
It is simplest to evaluate the eleven-dimensional equations of motion  
\ref{elevenDact} in the eleven-dimensional tangent frame. This
leads to the ten-dimensional equations
\begin{eqnarray}
\label{einstein}
R_{mn} - \frac{1}{2}g_{mn}R &=& - \frac{1}{2}e^{2\phi}\left(
F_{mp}F_{n}^{\ p} -\frac{1}{4}g_{mn}F^2\right)
-{1\over 4}\left(H_{mpq}H_{n}^{\ pq}- {1\over 6}g_{mn}H^2\right)\nonumber\\
&&
-{1\over 12}e^{2\phi}\left(G_{mpqr}G_n^{\ pqr}-{1\over 8}g_{mn}G^2\right)
+2\left(D_{m}D_{n}{\phi} - g_{mn}D^{2}{\phi} + 
g_{mn}(D\phi)^{2}\right) \nonumber\\
&& +18m(D_{(m}A_{n)} -g_{mn}D^{p}A_{p}) + 36m^{2}(A_{m}A_{n} +
4g_{mn}A^2)
\nonumber\\ 
&& +12mA_{(m}{\partial}_{n)}{\phi} 
+ 30mg_{mn}A^{p}{\partial}_{p}{\phi} + 144m^{2}g_{mn}e^{-2\phi}\ , 
\nonumber\\
\end{eqnarray}
\begin{eqnarray}
\label{scalar}
6D^{2}{\phi} - 8(D\phi)^2 = R &+& \frac{3}{4}e^{2\phi}F^2 +
\frac{1}{48}e^{2\phi}G^2- \frac{1}{12}H^2\nonumber\\
 &+& 360m^{2}e^{-2\phi}
+ 288m^{2}A^2 +96mA^{n}{\partial}_{n}{\phi} 
- 36mD^{n}A_n\ , 
\end{eqnarray}
\begin{eqnarray}
\label{maxwell}
D^{n}F_{mn} &=&-\frac{1}{6}G_{mnpq}H^{npq} 
+18mA^{n}F_{mn} + 72m^{2}e^{-2\phi}A_{m} -
24me^{-2\phi}{\partial}_{m}{\phi}\ ,
\end{eqnarray}
\begin{equation}
\label{twoform}
D^m(e^{-2\phi}H_{mnp}) = 12mA^mH_{mnp}e^{-2\phi}+\frac{1}{2}G_{npqr}F^{qr}
+{1\over 36\cdot 48}\epsilon_{npq...r...}G^{q...}G^{r...}\ ,
\end{equation}
\begin{equation}
\label{threeform}
D^mG_{mnpq} = 12m A^mG_{mnpq}+12mH_{npq}e^{-2\phi}
+{1\over 6\cdot 36}\epsilon_{npqr...s...}G^{r...}
H^{s...}\ ,
\end{equation}
where
\begin{eqnarray}
\label{gauge}
H_{mnp} &=& 3\partial_{[m}B_{np]}-6m C_{mnp}\ , \nonumber\\
G_{mnpq} &=& 4\partial_{[m}C_{npq]}+4A_{[m}H_{npq]}\ ,\nonumber\\
\end{eqnarray}
follow from the reduction of the eleven-dimensional four-form 
$\hat H_{abcd}$.
This reduction  agrees with the equations obtained in \cite{pope}
through a non-compact  `Scherk-Schwarz'
dimensional reduction of ordinary eleven-dimensional supergravity
(although these authors do not use the `string frame').

Note that the usual $U(1)$ symmetry of the gauge field $A_m$ is
broken. Since this is a result of eleven-dimensional diffeomorphisms
we expect that it is replaced by another symmetry. Indeed one can
show that these equations of motion are invariant under
\begin{eqnarray}
\label{sym}
\phi &\rightarrow& \phi -{3m}\chi\nonumber\\
A_m  &\rightarrow& A_m - \partial_m\chi\nonumber\\
g_{mn}  &\rightarrow& e^{-6m\chi}g_{mn}\nonumber\\
B_{mn}  &\rightarrow& B_{mn}+\partial_{[m}\Lambda_{n]}
+6m\Omega_{mn}\nonumber\\
C_{mnp} &\rightarrow& C_{mnp} +3\partial_{[m}\Omega_{np]}\nonumber\\
\end{eqnarray}
From the eleven-dimensional point of view these transformations
scale the metric by the Weyl factor $e^{-4m\chi}$.
Thus it is possible to gauge away $\phi$ and $B_{mn}$.

By construction these equations of motion are the restriction to 
Bosonic fields of a system which is invariant under the supersymmetry
\begin{eqnarray}
\label{susy}
\delta e_m^{\ \ \underline n} &=& -i\bar\epsilon e^{\phi/3}\Gamma^{\underline
  n}\psi_m - \frac{i}{2}\bar\epsilon e^{-2\phi/3}e_m^{\ \ \underline n}
\Gamma_{11}\lambda\ ,\nonumber\\
\delta\phi &=& -\frac{3i}{2}e^{-2\phi/3}
\bar\epsilon\Gamma_{11}\lambda\ ,\nonumber\\
\delta A_m &=&
-ie^{-2\phi/3}\bar\epsilon\Gamma_{11}\psi_m\ ,\nonumber \\
\delta B_{mn} &=& -i\bar\epsilon e^{-2\phi/3}\Gamma_{mn}\lambda
+2i\bar\epsilon e^{\phi/3} \Gamma_{[m|}\Gamma_{11}\psi_{|n]} 
+2i\bar\epsilon e^{4\phi/3}A_{[m}\psi_{n]}\ ,\nonumber\\
\delta C_{mnp} &=& -3i\bar\epsilon e^{-2\phi/3}\left(\Gamma_{[mn}\psi_{p]}
+\Gamma_{[mn}A_{p]}\lambda
-2e^\phi A_{[m}\Gamma_{n|}\Gamma_{11}\psi_{|p]}\right)\ , \nonumber\\
\delta \lambda &=& -{1\over3}e^\phi\partial^{n}\phi\Gamma_{11}\Gamma_{n}\epsilon
+{1\over8}e^{2\phi}F^{{mn}}\Gamma_{{mn}}\epsilon
-\frac{1}{36}e^{\phi}H^{{mnp}}\Gamma_{{mnp}}\epsilon
\nonumber\\ &&
+{1 \over 288}e^{2\phi}G^{{mnpq}}\Gamma_{11}
\Gamma_{{mnpq}}\epsilon+me^\phi A^{n}\Gamma_{11}
\Gamma_{n}\epsilon+2m\epsilon\ ,\nonumber\\
\delta \psi_m &=& D_m\epsilon+
\frac{1}{6}\partial^{n}\phi\Gamma_{mn} +{1\over 4}F_m^{\ n}\Gamma_{11}\Gamma_n
\epsilon+\frac{1}{36}e^{\phi}\Gamma^{npq}G_{mnpq}\epsilon \nonumber\\
&&
+\frac{1}{12}\Gamma^{np}\Gamma_{11}H_{mnp}\epsilon
+\frac{1}{288}e^\phi \Gamma_{mnpqr}G^{npqr}\epsilon +\frac{1}{72}\Gamma_{mnpq}\Gamma_{11}H^{npq}\epsilon
\nonumber\\
&&+ie^{2\phi/3}\bar\epsilon\Gamma_{11}\psi_m\lambda
+mA^n\Gamma_{mn}\epsilon + me^{-\phi}\Gamma_{11}\Gamma_m\epsilon
-2mA_m\epsilon\ ,\nonumber\\ 
\end{eqnarray}
where $\lambda= \hat \psi_y$ and 
$\psi_m = \hat\psi_m-A_m\lambda$ are the ten-dimensional 
dilatini and gravitini respectively, and
$\Gamma_{11} = \hat \Gamma_{\underline y}$. Although somewhat complicated
these supersymmetries are merely the symmetries of Weyl superspace
written in a ten-dimensional form (assuming no dependence on $y$)
and reduce to those of massless type IIA supergravity when $m=0$.

\section{Stability of de Sitter space}

The naive ground state of MM-theory, where the non-metric fields
are set to zero, is ten-dimensional de
Sitter space: $R_{mn}=-36m^2g_{mn}$. 
The first thing one notices about this de Sitter vacuum is that
the massive vector field $A_m$ is in fact tachyonic at the linearised
level.  Therefore one expects that de Sitter space is a ``false''
vacuum  of some
kind. To further understand this issue let us consider the
stability of de Sitter space under a small vector perturbation.
To this end we expand the equations of motion to lowest order
in $A_m={\cal O}(\epsilon)$, $g_{mn} = g_{mn}(dS) +{\cal
  O}(\epsilon)$,   
but with all the other fields vanishing.  This gives
\begin{eqnarray}
\label{linear}
R_{mn} - \frac{1}{2}g_{mn}R -144m^2g_{mn}&=& 18mD_{(m}A_{n)}
+{\cal O}(\epsilon^2)\nonumber\\
D^{n}F_{mn} &=&  72m^{2}A_{m} +{\cal O}(\epsilon^2)\ .\nonumber\\
\end{eqnarray}
Note that we have used the fact that the second equation implies 
$D^nA_n={\cal O}(\epsilon^2)$. This in turn
implies that \ref{scalar} is satisfied to lowest order and also that
the only non-vanishing term on the right hand side of \ref{einstein}
is the one given in \ref{linear}.
Taking the trace of the linearised Einstein equation shows that the background
is still of constant curvature: $R=-360m^2+{\cal O}(\epsilon^2)$. 
The second equation of \ref{linear} may now just be written
as
\begin{equation}
\label{linAeq}
D^2 A_n = -36m^2A_n+{\cal O}(\epsilon^2)\ .
\end{equation}

It is not hard to check that a natural family of solutions 
to these equations is just to let $A_m = \epsilon K_m$ where
$K^m$ is a Killing vector\footnote{Recall that, in general,
a Killing vector in a vacuum spacetime acts as a 
vector potential for a Maxwell test field \cite{wald}}. In this case we find 
$g_{mn} = g_{mn}(dS)+{\cal O}(\epsilon^2)$, i.e. to this order
the metric remains that of de Sitter space.
However if we choose coordinates where de Sitter space has the metric
\begin{equation}
\label{deSitter}
ds^2 = -dt^2 + e^{4mt}dx^idx^i \ ,
\end{equation}
where $i=1,...,9$, then a class of Killing vectors, corresponding to 
translations along $x^i$,
have the form $K^m = \delta^m_i$.
These produce the vector field $A_m = e^{4mt}\delta_m^i$. Thus a
small perturbation of de Sitter space at time $t=0$ by an electric
field will in fact
grow exponentially in size as space inflates until it is of order
one and we can no longer neglect the non-linear terms, including the
back-reaction on the metric. Thus we
conclude that the de Sitter space vacuum of MM-theory is unstable.

It is amusing to note that precisely this form of the tachyonic vector
field has been studied before on de Sitter space \cite{Turner}.
These authors showed that, in  an inflationary scenario, 
interactions for the tachyonic vector field give rise to
primordial magnetic fields of the same magnitude that
are observed in galaxies today. 
Here we see that such peculiar interactions readily appear in MM-theory.

We can also consider small perturbations about de Sitter space by
the two-form  and three-form fields. In particular consider a de Sitter
background and turn on a small $H_{mnp}$ and $G_{mnpq}$ of order
$\epsilon$. To order $\epsilon$ the metric remains that of de Sitter
space and the remaining equations of motion are
\begin{eqnarray}
\label{linep}
D^mG_{mnpq} &=& 12mH_{npq} + {\cal O}(\epsilon^2)\ ,\nonumber\\
D^mH_{mnp}&=&0 + {\cal O}(\epsilon^2) \ ,\nonumber\\
\end{eqnarray}
and now $G_{mnpq} = -\frac{4}{6m}\partial_{[m}H_{npq]}$. Clearly the
first equation implies the second. Therefore we simply need to consider
\begin{equation}
\label{linthree}
4D^mD_{[m}H_{npq]} = -72m^2 H_{npq}+ {\cal O}(\epsilon^2)\ .
\end{equation}
Thus the three-form also appears to be  tachyonic. 
In the coordinates \ref{deSitter}, \ref{linthree} splits into
two equations:
\begin{eqnarray}
\label{dual}
4\partial^k\partial_{[k}H_{0ij]} &=& -72m^2e^{4mt}H_{0ij}+ {\cal O}(\epsilon^2)\ ,\nonumber\\
4\partial_0\partial_{[0}H_{ijk]}&+&24m\partial_{[0}H_{ijk]} 
- 4e^{-4mt}\partial^l\partial_{[l}H_{ijk]}= 72m^2H_{ijk}+ {\cal O}(\epsilon^2)\ .
\end{eqnarray}
One simple solution to these equations is to
take $H_{ijk}=\epsilon e^{\lambda mt}C_{ijk}$, $H_{0ij}=0$ where $C_{ijk}$
is a constant tensor. Inserting this ansatz into \ref{dual}
leads to  $\lambda = 6$ or $\lambda=-12$. On the other hand one can
easily check that there are no solutions with $H_{ijk}=0$.
Thus, 
just as with the vector field above,  a small but constant 
three-form $H_{ijk}$ with $\lambda =6$ will
eventually grow so large that its effect on the de Sitter background
spacetime can no longer be ignored.

\section{Zero-branes}

Given that the de Sitter vacuum of MM-theory 
is unstable against the spontanous creation of electric fields, 
it seems natural to search for static solutions of the field 
equations that carry charges, i.e. solutions with 
$F_{0i}\ne 0$. To this end we start by  considering the 
familiar D0-brane ansatz
\begin{eqnarray}
\label{D0ansatz}
ds^2 &=& -e^{2\alpha\phi}dt^2 + e^{2\beta\phi}dx^idx^i\ , \nonumber \\
A_i&=&0\ .\nonumber \\
\end{eqnarray}
First we examine the non-linear vector equation \ref{maxwell}. 
Demanding that the solution is time-independent
leads to the relation 
\begin{equation}
\label{A0}
A_0 = \pm\sqrt{{4\over 3(1-\alpha)}}e^{(\alpha-1)\phi}\ .
\end{equation}
This then leaves the equation
\begin{equation}
\label{A0eq}
\partial_i\partial^i \phi +(7\beta-1)\partial_i\phi\partial^i\phi
=-{72m^2\over \alpha-1}e^{(2\beta-2)\phi}\ ,
\end{equation}
where the $\pm$ sign chooses between zero-branes and anti-zero-branes.
If we now subsitute this ansatz into the Einstein equation
\ref{einstein} we  find that $\alpha=-\beta=-1/3$. 
One can then check that 
\ref{scalar} is  automatically satisfied. Thus we find the solution is
simply
\begin{eqnarray}
\label{D0sol}
ds^2 &=& -H^{-{1\over2}}dt^2 + H^{1\over2}dx^idx^i\ , \nonumber \\
A_0 &=& \pm H^{-1}\ ,\nonumber\\
e^{4\phi/3} &=& H\ .\nonumber\\
\end{eqnarray}
This is precisely the same form as D0-branes in massless  type IIA string
theory. The only difference is that in MM-theory the function $H$ is
not harmonic but instead satisfies, from \ref{A0eq},
\begin{equation}
\label{source}
\partial_i\partial^i H = 72m^2\ .
\end{equation}
However one can check that this solution does not preserve 
any supersymmetries
since, although from the eleven-dimensional point of view it is `almost'
supersymmetric: $\delta\hat\psi_m=0$, $\delta \hat\psi_y=3m\epsilon$.

We note that $H$ can be written as 
\begin{equation}
\label{H0def}
H= H_0+4m^2A_{ij}x^ix^j \ ,
\end{equation}
where 
$A_{ij}$ is a constant matrix with ${\rm Tr}A=9$ and $H_0$
is a harmonic function  which we take to have the
usual $p$-brane form
\begin{equation}
\label{harmonic}
H_0 = 1 + \sum_n {Q_n\over |x^i-x^i_n|^7}\ . 
\end{equation}
Thus we can obtain solutions describing an
arbitrary collecton of zero-branes in a background ``condensate'' 
of electric charge. In particular
the near-horizon geometry of these zero-branes will be the same as
the D0-branes in the massless type IIA theory. 
On the other hand, as $r\rightarrow\infty$, where $r = x^ix^i$
the spacetime is not asymptotically flat.
Furthermore the dilaton grows without bound, so that asymptotically
the spacetime decompactifies.

The appearance of  $m^2$ in $H$ is reminiscent of
Reissner-Nordstrom-de Sitter solutions. However here the solution
is not asymptotically de Sitter and contains a non-vanishing dilaton. 
We may find another class of
solutions by invoking the $U(1)$ 
symmetry \ref{sym} of the equations of motion
resulting from eleven-dimensional diffeomorphisms.
In particular if we set $\chi = \phi/3m$ we can transform away the
dilaton and find the solution
\begin{eqnarray}
\label{newD0}
ds^2 &=& -H^{-2}dt^2 + H^{-1}dx^idx^i\nonumber\\
A_0 &=& \pm H^{-1}\ ,\quad A_i = -{1\over 4m}\partial_i {\rm ln}H \nonumber\\
\phi&=&0\nonumber\\
\end{eqnarray}
where $H = H_0+4m^2A_{ij}x^ix^j$ is the same as before.  Note that $F_{mn}$
is unchanged by this transformation, i.e. we still only have electric
fields. Although we note that now it is not clear how to take a 
smooth limit as $m\rightarrow 0$. However a smooth limit might exist 
since terms the divergent terms $A_i$ are pure gauge. Indeed 
the smooth limit  might require a transformation 
to the infinite momentum frame.

Let us consider the spherically symmetric case $A_{ij}=\delta_{ij}$.
Far from the zero-branes $H_0=1$ and hence  $H=1+4m^2r^2$. 
It is instructive to transform to
a new coordinate $\tilde r^{-2} = r^{-2}+4m^2$ in which case the
metric in \ref{newD0} becomes
\begin{equation}
\label{newnewD0}
ds^2 = -(1-4m^2\tilde r^2)^2dt^2 +{d\tilde r^2\over (1-4m^2\tilde r^2)^2}
+\tilde
r^2d\Omega_8^2\ .
\end{equation}
This is very similar to de Sitter space in static coordinates which has the 
metric
\begin{equation}
\label{deSittertwo}
ds_{dS}^2 = -(1-4m^2 r^2)dt^2 +{dr^2\over (1-4m^2 r^2)}
+ r^2d\Omega_8^2\ .
\end{equation}
Indeed \ref{newnewD0} 
has the same causal structure and in particular 
$\tilde r^2 = 1/4m^2$ corresponds to a
cosmological event horizon. Furthermore, at least in these
coordinates,  all of the curvature components are well behaved.
For distances much
less than the cosmological horizon the metric looks like that
of de Sitter (but with twice the cosmological constant of the
vacuum de Sitter solution). 

Lastly we note that the entropy of the zero-brane spacetime \ref{newnewD0},
viewed as one quarter of the area of the cosmological horizon, is 
\begin{equation}
\label{entropy}
S = {Vol(S^8)\over 1024 m^8}\ ,
\end{equation}
where $Vol(S^8)$ is the volume of a unit eight-sphere. 
This is the same as the entropy of the 
de Sitter vacuum. This is particularly
intriguing since this spacetime can contain arbitrarily many zero-branes,
each with eight bosonic degrees of freedom. Whereas
the entropy is supposed to measure the number of degrees of 
freedom in the corresponding quantum theory.\footnote{In
\cite{BFSS} it was argued that there should be an upper bound on
the transverse density of partons.  It would be interesting to
see if similar arguments may
imply that there is an upper bound on the total number of zero-branes 
which can be `packed' within the cosmological horizon.  
We will not pursue these
ideas here.}

We can also lift the solution to eleven-dimensions using the ansatz
\ref{veilbein}.
This leads to the standard pp-wave spacetime only with a non-harmonic
function $H$
\begin{equation}
\label{ppwave}
d{\hat s}^2 = Hdy^2 \pm 2dydt+ dx^idx^i\ . 
\end{equation}
Asymptotically, where we can neglect $H_0$, 
this spacetime is known as a Cahen-Wallach space. It can be constructed
as a smooth symmetric space  
and has also been recently studied in  conventional (i.e. $\hat k_a=0$) 
M-theory \cite{FP}.  A crucial difference, however, is that in \cite{FP}
the source for $H$ is provided by the four-form field strength, whereas
in our analysis the source is $\hat k$. In particular 
the four-form gives rise to a source for $H$ with
$m^2\rightarrow -m^2$,
i.e. anti-de Sitter, rather than de Sitter, behaviour.

Thus MM-theory naturally contains zero-branes and 
the cosmological constant provides a uniform and constant 
source for them. Since zero-branes are eleven-dimensional
gravitons carrying single units of momentum around the eleventh
dimension, this hints that MM-theory itself is unstable against
de-compactification to eleven dimensions, where it is physically
equivalent to M-theory. Thus we are led to the conjecture that
MM-theory is an unstable vacuum of M-theory. 

\subsection{Other Branes}

It seems natural to look for other static $p$-brane solutions with the
ansatz 
\begin{equation}
\label{pmetric}
ds^2 = e^{2\alpha\phi}\eta_{\mu\nu}dx^\mu dx^\nu 
+ e^{2\beta\phi}\delta_{ij}dx^i dx^j\ ,
\end{equation}
where $\mu,\nu=0,..,p$, $i,j=p+1,...,9$ and $\partial_\mu\phi=0$.
A first
attempt might be to look for six-branes, the magnetic duals
to the zero-branes. However if we set 
$G_{mnpq}=H_{mnp}=0$ and consider static solutions with $A_0=0$
then the vector equation \ref{maxwell} does not simplify in any
obvious way. Indeed the usual $D6$-brane solution is often
(but not necessarily) constructed by dualising $F_{mn}$ to
an eight-form. However the massive deformation discussed here makes
this dualisation non-local since $A_m$, and not just 
$F_{mn}$, appears in the equations of motion.

Given that the three-form $C_{mnp}$ also has a tachyonic instability
we might try to look for $D2$-branes and $D4$-branes. If we
now set $B_{mn}=A_m=0$ then \ref{maxwell} implies, for $m\ne 0$,
\begin{equation}
\label{Avanishing}
\frac{1}{6}\partial_{[m}C_{npq]}C^{npq} = e^{-2\phi}\partial_m\phi \ .
\end{equation}
For a $D2$-brane we expect that $C_{012}\ne 0$. This leads to
$C_{012} = \sqrt{8/(2-6\alpha)}e^{(3\alpha-1)\phi}$. 
However the Einstein and dilaton equation can only be satisfied if
$m=0$ and $\alpha=-\beta=-1$, in which case we obtain the 
standard D2-brane of the massless theory. 
In the case of four-branes we encounter the same problem that we saw for
six-branes, i.e. \ref{maxwell} does not simplify and it is not
clear how to dualise $G_{mnpq}$ since
the equations of motion involve $C_{mnp}$ without derivatives.

In particular it might seem as if there should be eight-brane
solutions, similar to those found in the Romans supergravity 
\cite{bergshoeff}. 
However inserting the ansatz \ref{pmetric} 
with $p =8$ into the equations of motion and setting 
$F_{mn}=H_{mnp}=G_{mnpq}=0$ ones finds that there are no 
{\it Poincare invariant} solutions.

\section{MM(atrix) Theory?}

The existence of solutions that represent an arbitrary number of
static zero-branes is analogous to the situation of  
D0-branes in massless type IIA
string theory. In addition it is well-appreciated that in a certain
limit, known as the infinite momentum frame,  the dynamics of D0-branes
contains the entire physics of string theory
and M-theory. In particular this allows for a matrix definition
of string theory and M-theory\cite{BFSS}. Therefore one might hope that some
modification of this matrix theory could be interpreted as the
dynamics of the zero-branes we found above in MM-theory. Furthermore this
naturally leads to a microscopic quantum definition of MM-theory, as a
modified (or massive) MM(atrix) theory.

To construct the effective action for the zero-branes we first recall
the analogous construction in of the D0-brane effective action starting from  
M-theory. Following \cite{paul} one starts with the superspace action
for a massless superparticle in eleven dimensions
\be\label{superp}
S = -\int d\tau{1\over \hat v}\hat E_\tau \cdot \hat E_\tau -\dot Y,
\ee
where $\hat E_\tau^{\ \underline A}$ is the pull back to the worldline of
the superveilbien $\hat E_{A}^{\ \underline B}$ 
and $\hat v$ is an independent worldline 
density. Here $Y$ is the  coordinate corresponding to the eleventh
dimension, and a dot denotes
differentiation with respect to the worldline coordinate $\tau$
\footnote{The last term is therefore a total derivative but is needed as
explained in \cite{paul}.}.
Using the analogous decomposition for the superveilbien
that we used in \ref{veilbein}, and assuming that nothing depends upon
$Y$, we can  
remove $Y$ and $\hat v$ using their equations of motion to obtain
(setting the Fermions to zero)
\begin{equation} 
\label{action}
S = -\int d\tau e^{-\phi}\sqrt{-g} - \dot X^m A_m 
\end{equation}
where $g =\dot X^m\dot X^n g_{mn}$ is the pull back of the 
spacetime metric to the
one-dimensional worldvolume, and $\dot X^m A_m$ is the Wess-Zumino term which
describes the coupling of the zero-brane to the RR field $A_m$. 

Let us now consider a superparticle in MM-theory.
If we imagine that $y$ is non-compact then we may map the
superparticle of M-theory to a superparticle of MM-theory by 
rescaling $E_A^{\ \underline B}\rightarrow 
e^{-2\hat\theta}E_A^{\ \underline B}$ where  $\hat k=d\hat\theta$.
However in \ref{superp} this  transformation is simply
absorbed by the worldline density $\hat v$. Clearly this construction
is unaffected if we now compactify $Y$, assuming that ${\hat v}$ is
again taken to be independent of $Y$.
Thus we obtain precisely
the same worldline action \ref{action} for the zero-branes in MM-theory.
As a check on this construction we note that \ref{action} is invariant
under the MM-theory gauge transformation \ref{sym}.

Let us consider two backgrounds for these zero-branes. The first example 
is de Sitter space and we choose the coordinates \ref{deSittertwo}. 
Using static 
gauge $\tau =t$ and keeping only the 
radial coordinate $r$, the zero-brane action in 
this background is
\begin{equation}
\label{actone}
S = -\int dt \sqrt{(1-4m^2r^2)-{\dot r^2\over 1-4m^2r^2}}
=\int dt {1\over 2}{\dot r^2\over (1-4m^2r^2)^{3/2}} 
-\sqrt{1-4m^2r^2}+\ldots\ ,
\end{equation}
where we have kept only the lowest order term in a derivative expansion. 
Here we see that de Sitter space induces a potential on the worldline
action of the zero-branes. Near $r=0$ this potential appears tacyhonic.
Secondly we consider a zero-brane in 
the background of other zero-branes. Using static gauge the action becomes
\begin{equation}
\label{acttwo}
S = -\int dt H^{-1}\sqrt{1-H\dot x^i\dot x^i}-H^{-1}
=\int dt {1\over 2}\dot x^i \dot x^i+\ldots \ , 
\end{equation}
where again we have only kept the lowest order terms in a derivative expansion.
Here we see that at lowest order terms are free and identical to those
of D0-branes in type IIA string theory.

Next we must consider the effective action for $N$ zero-branes. 
In M(atrix) theory one considers the light-like limit and only
the lowest order terms in the effective action survive. Furthermore
one knows from D-brane quantisation via open strings that for $N$ D0-branes
these terms are precisely that of  a maximally supersymmetric
one-dimensional $U(N)$ gauge theory
\be\label{SQM} 
S_{M(atrix)} = \int d\tau{\rm Tr}\left[ {1\over2}\dot X^i\dot X^i-{1\over 2}
\sum_{i<j}[X^i,X^j]^2\right]\ .
\ee
The free action \ref{acttwo} is recovered from \ref{SQM} by 
identifying $x^i$ with the $U(1)$ part
of $X^i$.
Therefore our next task is to motivate a proposed modification 
to the quantum mechanics if $m\ne0$. 

There is at least one crucial difference between any proposed 
matrix definition of MM-theory
theory and M(atrix) theory. Namely in M(atrix) theory the near horizon
geometry of the D0-branes can be obtained through a limit of  M-theory
where the compact $S^1$ becomes light-like. To see this one starts
with the usual D0-brane metric of type IIA string theory
(i.e. \ref{D0sol} but with $m=0$) and  sets $H =1+h$, 
where $h$ vanishes at infinity. 
In the massless theory we can choose the graviphoton to have 
the form $A_0=H^{-1}-1$, which differs from 
the zero-brane solution \ref{D0sol} 
by a gauge transformation. In this case one finds that the
eleven-dimensional lift of this solution does not take the form \ref{ppwave}
but rather
\be
\label{lightlight}
ds_{11}^2 = -dt^2 + dy^2 +hd(y-t)^2+dx^idx^i \ .
\ee
Reducing to ten-dimensions on the (asymptotically)
light-like circle $y-t$ then 
leads to the near-horizon geometry of the D0-branes (i.e. one finds
\ref{D0sol} but with $H$ replaced by $h$). The interpretation of
this observation is that M(atrix) theory describes M-theory states
that have been asymptotically boosted into the
infinite momentum frame. 
However in MM-theory we cannot so simply choose the graviphoton to have
this  form
since the required gauge transformation also induces a Weyl rescaling
of the metric and a shift in the dilaton. 
Instead if we set $\chi=t$, so that $A_0=H^{-1}-1$,  we find 
\be
\label{lightliketwo}
ds_{11}^2 = e^{-4mt}\left[  -dt^2 + dy^2 +hd(y-t)^2+dx^idx^i\right]\ .
\ee
Compactification on $y-t$ again
reproduces the near horizon geometry of the zero-branes (in the gauge
transformed variables). However asympotically $h$ now diverges
so that $y-t$ is never light-like. 
Of course it is not surprising that the near horizon geometry of
these zero-branes cannot be related 
to asymptotically light-like compactified
MM-theory, because MM-theory does not have asymptotically flat solutions.

Indeed, since the $U(1)$ symmetry \ref{sym} lifts to a diffeomorphism 
combined with a Weyl transformation of the eleven-dimensional metric, 
we see that by a choice of gauge we can arrange for an
eleven-dimensional solution which  of the form \ref{lightliketwo}
but with any conformal factor $e^{-4m\chi}$. Asymptotically 
$H=1+h \sim 4m^2A_{ij}x^ix^j$ so that from the eleven-dimensional
point of view  the zero-branes describe states in MM-theory which are
asymptotically conformally Cahen-Wallach.

This conformal symmetry implies that there is no suitable 
Maldacena limit. In particular, from the point of
view of MM-theory it is not meaningful to consider the limit where
the energy and length scale of the system is taken to zero, since
this requires a choice of conformal frame (or, from the
ten-dimensional point of view, a choice of gauge for the graviphoton). 
In effect the zero-brane dynamics are not associated
with the  light-like limit of MM-theory but rather 
they somehow describe MM-theory at all scales. As a result it seems
likely that we can not restrict our attention the lowest order
terms in zero-brane action \ref{action} which then leads to \ref{SQM}
if there are $N$ zero-branes. We therefore imagine a non-trivial 
phase of matrix quantum mechanics where none of the higher
order terms can be neglected. This resolves the puzzle that at lowest
order in fields, the zero-brane action \ref{action} does not depend on $m$
and so one can not tell the difference between MM-theory and M-theory.
However this is not the case if the higher order terms are not neglected.
In addition the tachyonic nature of zero-branes in a de Sitter
background is stabilised through higher order terms. 

It is intriging to note that Cahen-Wallach spacetimes are symmetric spaces 
with a  coset 
structure $G/H$, where $G$ is the Heisenberg algebra associated to
a system with nine degrees of freedom, combined with an
outer automorphism that intertwines the momentum and coordinate
variables \cite{FP}. Specifically, $G$ is
the algebra whose generators statisfy \cite{FP}
\begin{equation}
\label{algebra}
[p_i,q_j]=4m^2A_{ij}\ ,\quad [R,q_i]=p_i \ ,\quad [R,p_i]=4m^2A_{ij}q_j\ ,
\end{equation}
with all other commutators vanishing. The subalgebra 
$H$ is generated by the momenta $p_i$ and the corresponding symmetry is
presumably that of coordinate translations.
We are led to
conjecture that $G/H$ is the symmetry group of the underlying quantum
mechanical MM(atrix) theory in an asymptotically conformally Cahen-Wallach
spacetime. Since $H$ is linearly realised this suggests that there is no 
potential in the Hamiltonian.
It would be interesting to see if
the generalised conformal mechanics of D0-branes \cite{jevicki}
is related to a possible MM(atrix) theory.

This situation can be sharply contrasted with the D0-branes of the
massless IIA string.  In that case, the supergravity solution 
is asymptotically flat and one may choose the 
string coupling to be arbitrarily small so that
ordinarily the physics would be strictly ten-dimensional.  However, if one
stacks a large number `$N$' of D0-branes in the core of the spacetime, then a 
bubble of eleven-dimensional spacetime will open up in the region
$N^{1/9}/M < r < N^{1/7}/M$ \cite{polch,rajesh}.  For the massive
zero branes that we have found this effect will persist, the key difference
is that the geometry can never be asymptotically flat. 

Given that the symmetry which we use to (locally) transform from
M-theory to MM-theory is also a symmetry of the D0-brane effective action,
there is a bit of a puzzle here: How could
de Sitter space possibly be the ground state of MM-theory, given that it is
definitely {\it not} a solution of the massless IIA string?  
Again, our main answer to this conundrum is that there should exist
a phase of matrix quantum mechanics where higher order terms survive.
However, it is possible to see that this theory will be sensitive
to the fact that the bulk R-R vector is tachyonic, even at the
linearized level.  This is because
on a D0-brane the world-volume 
quantum mechanical theory includes sixteen Fermionic operators
$\theta$ \cite{halpern}.  The zero-modes of these Fermions generate an $SO(16)$ 
Clifford algebra,  and consequently may
be written as $2^{16/2} = 256$-dimensional gamma 
matrices.  It follows that a D0-brane has
256 internal degrees of freedom, or polarization states.  These states 
are explicitly
constructed in~\cite{mark}, and they are precisely the space of
polarization states of the supergraviton in eleven dimensions.  
In the weak field
approximation, the worldvolume fermions couple to small fluctuations
of the background metric, $g_{mn}$, NS-NS 2-form potential, $B_{mn}$, 
and R-R one-form and three-form
potentials, $A_m$ and $A_{mnp}$ \cite{morales, mark}.  The precise 
interaction Lagrangian for the R-R vector takes the form
\begin{equation}
L_{int} = -\frac{i}{8} \left({\nabla}_{m} {A_n} \right) \bar{\theta} 
{\Gamma}^{mn} {\theta}\ .
\end{equation}
In this sense the zero-brane internal degrees of freedom are `sensitive' to
an MM-theory background.  Understanding this interaction properly will involve
working out the full family of superpartners for the purely bosonic zero-brane
which we have been discussing in this paper.  In other words,
while a finite $N$ matrix formulation of DLCQ MM-theory should provide 
a (partial) microscopic description of MM-theory, it would also be 
of interest to 
better understand the asymptotic structure of these massive zero-branes.
In particular, it would be interesting to work out the long-range supergravity
fields which arise when we polarize one of these massive Bosonic zero-branes
with a higher multipole moment.  For example, it should be possible 
to polarize one of these zero-branes with a two-brane dipole moment  
as discussed 
in \cite{Wati}. However, this is complicated by the observation above that
it seems difficult (though not necessarily impossible) to find a two-brane
solution for this theory.

\sect{Relaxation of the cosmological constant through domain wall nucleation?}

As described above, in order to construct the new massive IIA 
supergravity of \cite{neil}, we must introduce an exact one form 
$\hat k = mdy$ in eleven-dimensions.
As described in \cite{andrew}, 
in ten-dimensions we can introduce a ten-form field strength 
\be
\label{tenform}
F_{10} = \star m\ .
\ee
Thus $d\star F_{10}=0$ and $dF_{10}=0$ automatically. 
From this point of view it is natural to consider the existence of
eight-brane or domain wall solutions that couple to $F_{10}$.  Of course,
when we say that states of this massive IIA theory couple electrically
to the 10-form $F_{10}$, what we really mean is that there exists a nine-form
potential $A_9$, which couples to the worldvolume of the eight-brane, 
and which is related to the ten-form in the usual way:
\be
\label{nineform}
F_{10} = dA_{9}\ .
\ee
Unfortunately, at present there is no reason why we should make 
this assumption.
This is related to the fact that at present we still don't have a good
microscopic realization of the degrees of freedom underlying the ten-form
formulation of this theory.  This can be contrasted with the Romans
massive IIA theory \cite{roman}, where the D8-branes are of course places
where F-strings can end.
We have also seen that no Poincare invariant eight-brane solutions to the
equations of motion exist, whereas they do exist in the Romans theory 
\cite{bergshoeff}. 

At any rate, if we could justify the statement that such domain walls  exist
then we could invoke the mechanism of 
Brown and Teitelboim (\cite{brown87}, \cite{brown88}),
who originally studied the stability of a theory in four dimensions where the
effective (positive) cosmological constant is generated by a four-form
flux, with 
fundamental membranes coupled electrically to the four-form.  They proved
that the effective cosmological constant will be decreased 
through the nucleation of membranes which couple to the four-form.
This decrease of the cosmological constant through brane nucleation is
generically known as the {\it Brown-Teitelboim mechanism}.
Clearly, if the massive IIA theory which we have been studying is just
a higher-dimensional realization of the scenario originally studied by
Brown and Teitelboim, then we may use the Brown-Teitelboim
mechanism to conclude that at late times the cosmological constant will be 
driven to zero.  Research on this possibility is currently underway.

\section{Conclusions}

In this paper we have addressed some dynamical issues in a modified
type IIA supergravity. In particular we showed that it admits
multi-zero-brane solutions and that the effective desciption of
these zero-branes can be related to a massive deformation of the
well-known M(atrix) formulations of M-theory and type IIA string theory.
This suggests a matrix theory formulation of MM-theory in which 
the massive equations of motion arise as the
the low energy effective dynamics.

The massive deformation of type IIA supergravity discussed here 
can almost certainly be extended to other supergravities. Indeed
the mechanism is in some sense trivial. What is, in our opinion,
non-trivial is that this construction is completely compatible with
eleven-dimensional superspace. Indeed it is the most general solution
to the constraints of eleven-dimensional supersymmetry. We find that
this is a compelling reason to explore the conjecture that it 
also has a microscopic description related to that which
underlies M-theory.  We find this motivation particularly
exciting, since our results imply that a cosmological 
constant is generated spontaneously.

Extrapolating the discussion here of M-theory to a realistic
four-dimensional phenomenology, we are presented with the following
inflationary scenario: at its creation the universe is in fact 
three-dimensional. It undergoes
a period of very rapid, Planck scale, inflation where an
additional fourth-dimension opens up. The size of this
dimension also rapidly expands well beyond any meaningful 
observable scale. The eventual ground
state in such a model is then four-dimensional Minkowski space (perhaps
with a small remnant of the cosmological constant).
The idea that our universe is, in some sense, three-dimensional
has appeared before. Notably with Witten's suggestion that the
cosmlogical constant problem might be solved if our universe
is in a strongly coupled phase of a three-dimensional supersymmetric
theory \cite{witten3D}.  It would be intriguing if the supergravity
discussed here provides a cosmological realization of this mechanism.

\hbox{}

{\noindent \bf Acknowledgements}\\

The authors thank S. Chaudhuri, P. Howe, C. Hull, D. Minic, R. Myers,
G. Papadopoulos, I. Sachs  and E. Witten for discussions and comments 
on an earlier draft of this paper. AC was supported at MIT in part 
by funds provided by the U.S. Department of Energy (D.O.E.) under
cooperative research agreement DE-FC02-94ER40818, and is supported
by a Director's Funded Fellowship at Los Alamos National Lab.
N.D.L. is supported by a PPARC fellowship and special grant
PPA/G/S/1998/0061. N.D.L. would also
like to thank the Physics departments of the Univeristy of Pennsylvania and
the Ludwig-Maximilians Universitiet where some of this work was completed.

\section*{Appendix: Conventions}

Here we give the conventions we have used for the spin connection and
curvature. We follow the same conventions as in \cite{neil} where
underlined indices refer to the tangent frame, $m,n=0,...9$ label 
ten-dimensional indicies and
$a,b=0,...,10$ eleven-dimensional ones.  We also use hats to
denote eleven-dimensional fields. When discussing $p$-branes the
worldvolume coordinates are labelled by $\mu,\nu=0,...,p$
and the transverse coordinates by $i,j,k,...$.

We define the spin connection by
\begin{equation}
\label{A1}
\partial_m e_n^{\ \ \underline p} - \partial_n e_n^{\ \ \underline p} 
+\omega_{m\underline q}^{\ \ \ \underline p}e_n^{\ \ \underline q}
-\omega_{n\underline q}^{\ \ \ \underline p}e_m^{\ \ \underline q}
 =0\ ,
\end{equation}
and hence the torsion-free Christoffel components can be obtained
as
\begin{equation}
\label{A2}
\Gamma_{mn}^p = e_{\underline q}^{\ \ p}
(\partial_m e_n^{\ \ \underline q}
+\omega_{m\underline r}^{\ \ \ \underline q}
e_{n}^{\ \ \underline r})\ .
\end{equation}
The curvature tensor is defined as
\begin{equation}
\label{A3}
R_{mnp}^{\ \ \ \ \ q} = \partial_m\Gamma_{np}^q -
\partial_n\Gamma_{mp}^q
-\Gamma^r_{mp}\Gamma^q_{rn} +\Gamma^r_{np}\Gamma^q_{rm} 
\ ,
\end{equation}
and the Ricci tensor is $R_{mn}= R_{mpn}^{\ \ \ \ \ p}$.
In particular one finds that $[D_m,D_n]V_p = -R_{mnp}^{\ \ \ \ \ q}V_q$.

\end{document}